# Vulnerabilities and Attacks Targeting Social Networks and Industrial Control Systems

Dharmendra Singh[1], Rakhi Sinha[2] , Pawan Songara[3] and Dr. Rakesh Rathi[4]

[1]Department of Computer Engineering, Govt. Engineering College, Ajmer
dharmendra.singh234@gmail.com
[2]Department of Information Technology, Govt. Engineering College, Ajmer
srakhi16@yahoo.com
[3]Aricent Group Of Industries, Gurgaon
pawansongara@aricent.com
[4]Department of Computer Engineering, Govt. Engineering College, Ajmer
rakeshrathi4@rediffmail.com

*Abstract:*

*Vulnerability is a weakness, shortcoming or flaw in the system or network infrastructure which can be used by an attacker to harm the system, disrupt its normal operation and use it for his financial, competitive or other motives or just for cyber escapades.*

*In this paper, we re-examined the various types of attacks on industrial control systems as well as on social networking users. We have listed which all vulnerabilities were exploited for executing these attacks and their effects on these systems and social networks. The focus will be mainly on the vulnerabilities that are used in OSNs as the convertors which convert the social network into antisocial network and these networks can be further used for the network attacks on the users associated with the victim user whereby creating a consecutive chain of attacks on increasing number of social networking users. Another type of attack, Stuxnet Attack which was originally designed to attack Iran's nuclear facilities is also discussed here which harms the system it controls by changing the code in that target system. The Stuxnet worm is a very treacherous and hazardous means of attack and is the first of its kind as it allows the attacker to manipulate real-time equipment*.

*Keywords:*

Online social networks, JPGvirus, Stuxnet worm, reverse social engineering attacks, Facebook application.

## 1. INTRODUCTION

The popularity of online social networks (OSN) [1] is vastly growing in today's world. The online communities developed by the OSNs are growing environments on the web to support modern interaction among the people around the Globe. Online social networks are very beneficial for the people for keeping in touch with friends, family members and other acquaintances, for research collaboration, data sharing, social campaigning and other constructive purposes. There are various social networks which are put to different purposes according to their usability. Some are used for professional and business collaborations like LinkedIn [2] and XING [3], whereas there are other OSNs like Facebook [4], Twitter [5], MySpace [6] and Orkut [7] which are primarily used for informal friendly interactions, photo and video sharing and entertainment.

The massive acceptance of OSNs by the users provides opportunity to the attackers to create and launch new exploit and attacks every day. The growing popularity of facebook in turn, makes it a popular website vulnerable to reverse engineering attacks [8] and identity thefts and may turn a social network into an antisocial network. Antisocial network is the platform for the execution of malicious threats as well as it provides unauthorized accesses like Denial of Service attack, propagation of malware etc. A very large distributed database is used for OSNs and this acts as an advantage to make exploitation ideal because OSNs include community of users that share the same interests in the form of applications. These users share the same applications with the help of platform openness. This openness results in a user installing the application which is malicious and infected. These above characteristics give attackers the chance to manipulate that user to act as antisocial against all the internet users which are connected to the victim user. In upcoming sections of this paper, the properties of Facebook which is a Real World Social Network are discussed and then the study about utilization of these properties to launch attacks on facebook users is explained.

This survey paper also discusses about Stuxnet [9] which is an intelligently designed worm and uses some vulnerability in print spoolers and removable drives to make its way through windows based systems but ultimately targets industrial control systems. Industrial Control systems (ICS) used for a variety of industrial processes like manufacturing and refining, power generation, water management etc. [14] This type of critical infrastructure is managed by cyber-physical systems like SCADA [15] and use of internet based technologies to manage SCADA systems make them more susceptible to network based attacks with the intention to modify the code structure of PLCs (Programmable Logic Controllers).

## 2. BACKGROUND:

### A. Social Network

Large number of people in the world use social networking sites like Facebook, LinkedIn in their daily life. Facebook is the fastest social networking site which is surfed very much across the globe [10]. Developers of the Facebook developed a platform and a large collection of applications for this website to make it more effective and interesting to the naive Facebook users.

Users use these applications on facebook and share them according to common interests in their groups or communities and also invite their friends to share the application. Many applications are malicious and harmful among these applications which can affect the users' personal information and also affect their friends in the same facebook community.

### B. Puppetnets

Puppetnet [11] is an application which is used for the exploitation purpose of World Wide Web. There are different types of links included in the web page which can include many malicious links in between that can be used to harm the users on social networking sites like facebook, LinkedIn etc. Adversary or malicious user can also make special pages that include many malicious attacks and if the user downloads the malicious data from these links or pages, then the adversary can get benefit out of it.

### C. Statistical Distribution of infected machines by Stuxnet Worm in different countries of the world:

When the Stuxnet worm was created, figures pointed out high rate of infection at some specific sites. But when the worm's behaviour was discovered, there was a notable decrease in number of infections in the same areas. The following pie chart [16] and table [16] shows percentage of infected systems by Stuxnet worm since 2010 for top 14 countries. The most affected country is Iran followed by mostly Asian countries. United States has only a small percentage of this infection. The high infection rate in some regions and low rate in the others depends on many factors, one of which is good quality security software is being used or not that is, in Asian countries most people like to use the free or trial versions of antivirus software that do not provide maximum protection against malware, malicious websites and they do not provide robust firewalls and shields.

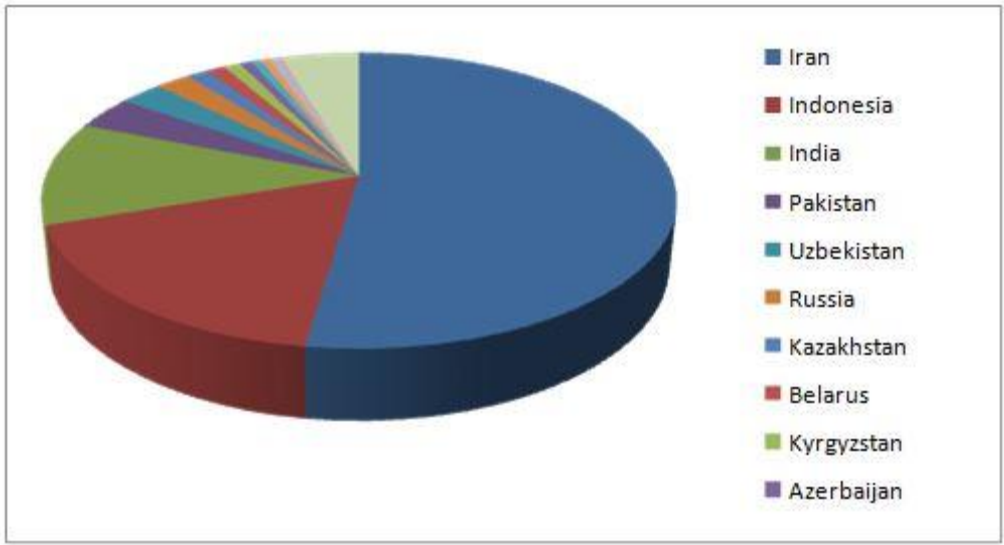

**Figure1:** Pie Chart depicting the percentage spread of Stuxnet Worm infection across the globe

**Table 1:** Percentage of Stuxnet Worm infection across the globe

| Iran | Indonesia | India | Pakistan | Uzbekistan | Russia | Kazakhstan | Belarus |
|---|---|---|---|---|---|---|---|
| 52.2% | 17.4% | 11.3% | 3.6% | 2.6% | 2.1% | 1.3% | 1.1% |

| Kyrgyzstan | Azerbaijan | United states | Cuba | Tajikistan | Afghanistan | Rest of the world |
|---|---|---|---|---|---|---|
| 1.0% | 0.7% | 0.6% | 0.6% | 0.5% | 0.3% | 4.6% |

**D. Phases to create and host a facebook application:**

Facebook, a very popular social networking website includes a provision which allows the software developers around the globe to develop simple applications which then execute along with facebook network and can access user data such as friends' list, inbox etc [12].

To create a facebook application, a developer application is required to be added to the facebook account of the developer and the development environment can be chosen from the various facebook compatible environments like PHP, JavaScript, Flash/Action Script and Connect. The facebook non-compliant environments are Android, ASP.NET, ASP (VB Script, Jscript ), Cocoa, Cold- Fusion, C++, C#, D, Emacs Lisp, Erlang, Google Web Toolkit, Java, Lisp, Perl, Python, Ruby on Rails, Smalltalk,TcI, VB.NET, and Windows Mobile. The client libraries of the above frameworks are not yet supported by facebook but are likely to be added to facebook client libraries in future.

The three essentials to be performed to implement and submit a facebook application are:

i. A third-party web server where the hosted application will reside.

ii. A facebook client library that is installed on hosting server.

iii. A relational database management solution like MySQL for storing application user database.

**a. Registering a facebook application:**

A developer registers his creation in the form of facebook application by submitting a form specifying the details like application name and description, Canvas page URL and Canvas callback URL. A canvas page URL is the introductory page of a facebook application to which a user is redirected when he chooses to open a particular application whereas; the canvas callback URL is the location of the application on its host server. The developer then receives a confirmation from the facebook development team if the application is selected and included in the global facebook application directory from where the facebook users can now explore and add the application to their account.

Some facebook elements that come handy to serve the purpose of easy application development are:

i. *Facebook Markup Languag (FBML):* FBML as the name suggests, is derived from HTML and includes tags with following syntax: <fb: tagName/>. With the help of FBML, the application communicates with user profile to request user data.
ii. *Facebook Query Language (FQL):* Various tables have been supplied by facebook to be queried for social user data.
iii. *Facebook JavaScript (FBJS):* FBJS allows facebook application developers to use JavaScript, AJAX objects, animation libraries and other client libraries to develop versatile facebook applications.
iv. *Facebook API:* APIs provide HTTP based communication with facebook API REST server and enables the developer to gather a particular user's profile information.

**b. FBML canvas page redirection mechanism:**

The FBML canvas page redirection mechanism flow is shown in the above figure[17]. Firstly, a facebook user requests an application from the facebook server. In order to redirect the user's browser to the application's canvas URL, Facebook server asks for the call-back URL from the hosting server through an HTTP post request for retrieving the FBML of the canvas page. If some data is needed by the web hosting server, then it also sends an HTTP GET request to the Facebook REST server. After completion of these API method calls, the hosting server creates and returns the FBML of the requested application page to the Facebook server. The facebook server then translates this FBML into HTML and provides it to the user's browser.

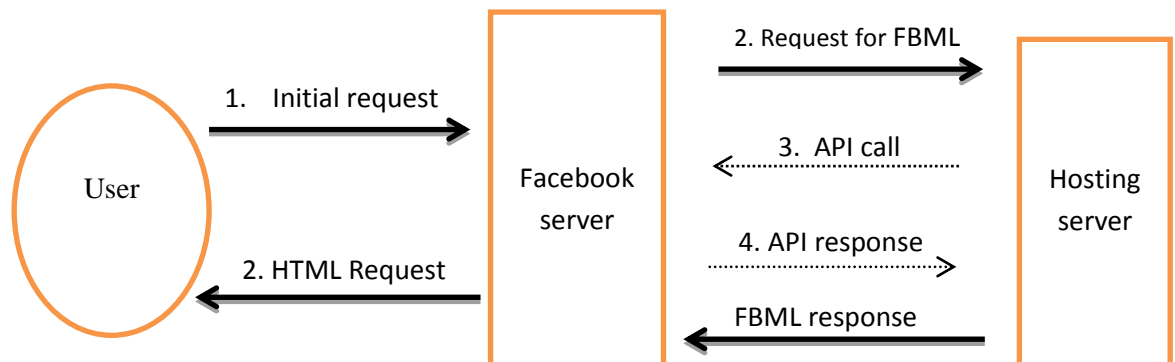

**Figure2:** How an application page is displayed on a user's browser

### E. Phases of Stuxnet:

Stuxnet is the most sophisticated and dangerous worm that was introduced in July 2010, by attacking Iran's nuclear facilities. Stuxnet affects approximately 60% of that attack on Iran as reported by Director of security response. According to the study done by Security experts in Iran, Stuxnet basically affects the Uranium facility due to which its rotational speed increases and decreases and as a result violates its normal behaviour.

Stuxnet worm basically works in two phases: In first phase that is the "**Propagation phase**", it propagates through the files and updates them. After that, in the second phase that is called the "**Injection phase**", it searches for its actual target that is i-e Siemens WinCC control and monitoring systems. After finding them, it starts to deviate them from their normal behaviour. We have introduced those three vulnerabilities above that can be simulated through Metasploit framework. This framework can also be used to write code for different types of vulnerabilities. The following Figure[16] shows these phases:

The Stuxnet rootkit spreads itself in four ways: with the help of flash drives or USB drives, while sharing data on networks or through print spooler or RPC vulnerability.

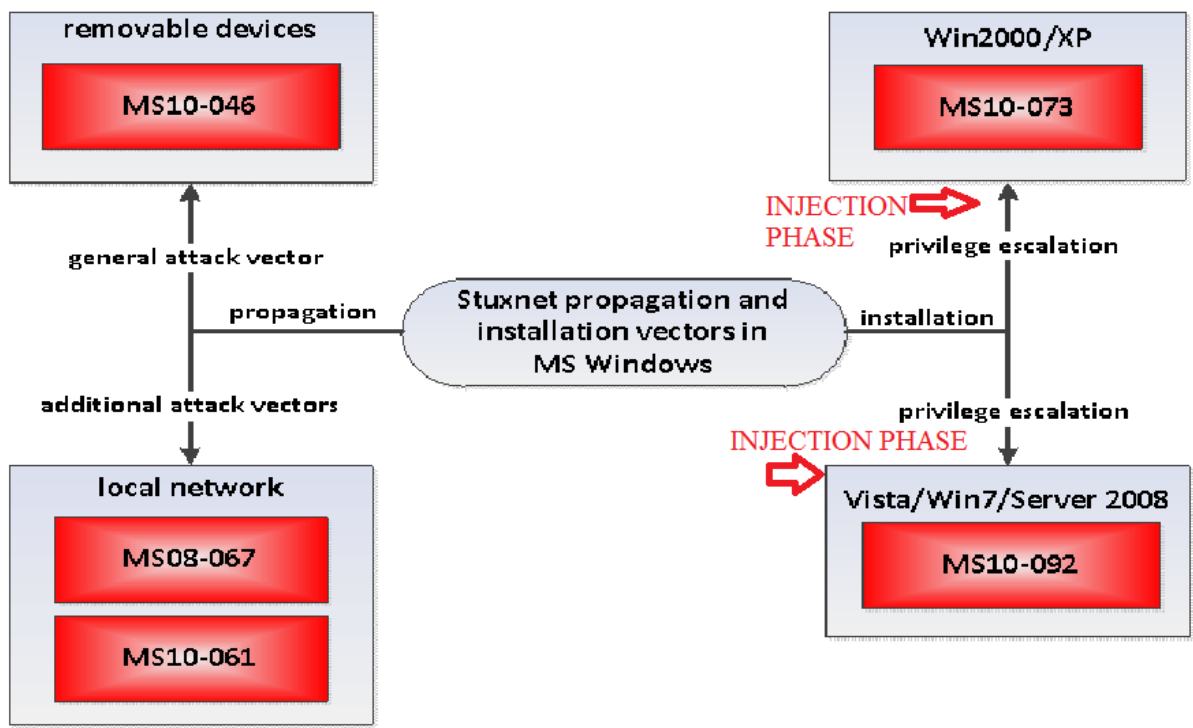

**Figure 3:** Two Phases of Stuxnet Worm

## 3. Experimental Evaluation

### 3.1 Experimental Evaluation for facebook:

#### 3.1.1 OSN as DDoS Platform
Social websites like Facebook can be exploited or infected through Distributed Denial of Service attack [13] by any malicious application which generates self-executable files on victim's system. When a user works on the system or opens a file, image etc, which has the hidden self-executable files, these files are spread in the whole system. The collection of computers that generate harmful files upon running malicious application on Facebook is called JPGvirus.

### 3.1.2 Experimental Setup

This consists of creating a JPGvirus but it does not affect any facebook user. A common application of facebook is Photo album creation. When a facebook user clicks on a photo, it gets opened. The attack is preceded in the following way: Special FBML tags are inserted into the code, so whenever a user visits that page, a self-executable file named windows.exe is created inside Windows folder in C Drive of PC.

A simple C program is created in the desired directory. Its executable is generated and is attached to image file with JPG extension. FBML tag <fb:iframe/> is used to include the hidden frames in HTML's code. A technique which is discussed later is used to protect the privacy of a facebook user and prevent the extraction of information from the image requests that are made by user's system compromised by malicious applications.

Possible misuses in the fashion of Puppetnets:

I. *Host Scanning*
In this technique, hosts are scanned by the attacker for finding out an open port. Whenever the attacker finds an open port, he will send an HTTP request.

II. *Embedded Self-Signed Java Applets*
The common idea used in this is that java applets are embedded inside the developer's applications and use this technique to create malicious self-signed applets which initiate the attack by asking the user to upload a file through a common file selector on their system. When the user uploads some file, then that self-signed applet has full control over the disk and accesses the user's local disk.

III. *Personal Information Leakage*
The facebook enables the users to have full control over the information provided by it. User can also apply some privacy settings to keep their data private but the attacker can get the personal information provided by the user and use them to create fake profiles or access their accounts by using their personal information if the users install the malicious applications created by attacker.

IV. *Malware Propagation*
Malware can be propagated by the facebook applications if the server is exploited by URL embedded attack vector. When the user uses this application which is infected, this malware will be propagated.

V. *URL Scanner Cloaking*
In this technique, requests coming from some specific users can be filtered by facebook URL scanning system and the attacker then feed the customers with unwanted information or data. Content forging is yet another method of attacking in facebook.

### 3.2 Experimental evaluation for Stuxnet:

All of the three vulnerabilities caused by Stuxnet worm are defined below. This can be understood by an architectural diagram in which C&C server behaves as Backtrack 4 and Metasploit Framework acts as USB:

### 3.2.1 MS08_067_netapi (Server Service)

This type of vulnerability is caused by Stuxnet by spreading itself through network either by file sharing or through shared folders. The PC which has been infected through this worm is connected to network and has the shared folder. When the other system on the network wants to access that shared folder, this vulnerability exploits that system also, through this shared folder or file.

### 3.2.2 MS10_061_spoolss (Print Spooler Server)
The Print Spooler vulnerability caused by Stuxnet worm is MS10-061. This worm affects the systems that share a common printer. The effects due to this worm are done in basically two steps: injection phase and execution phase. Injection phase involves copying infected files into directory and execution phase does the main work of infection by impersonating a client, for instance, by sending two files in place of a single file.

### 3.2.3 MS10_046_dllloader (.LNK Vulnerability)
When C&C server gives some specific commands, the coding of some system that is connected to it, gets changed. Then, at the time of execution, it will start giving alerts. Metasploit framework is an important tool that is commonly used to exploit different windows vulnerabilities. It is the first way in which Stuxnet rootkit distributes itself to other systems and infects them. According to the given architecture, this vulnerability is exploited to system containing Keil and proteous. Due to C&C server, the coding of system is changed and then the final output of that system will also be changed.
It basically loads the icons of files from CPL Windows control panel and changes the file location info field in .LNK header. In this way, attacker finally succeeds in gaining the access to sensitive information when user proceeds to that malicious path.

The above discussed three vulnerabilities can be simulated by using Metasploit Framework integrated with Linux Backtrack 4 in the below figure[9]. Basically, security professionals and researchers use Metasploit framework to exploit vulnerabilities, perform penetration testing. Backtrack is an extension of Linux which is particularly designed for performing functions aimed at maintaining network security for security professionals.

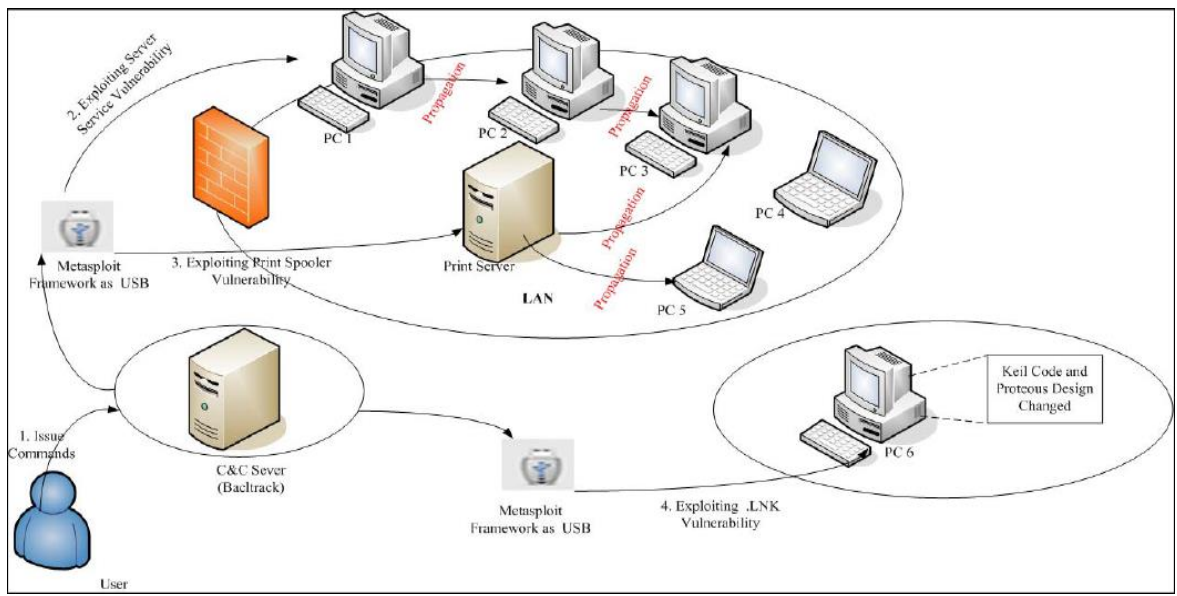

**Figure 4:** Network Setup for the Transmission of Stuxnet Worm to the Target System

### 3.3 Defenses against JPGvirus and Stuxnet vulnerabilities

The JPGvirus and Stuxnet vulnerabilities explained in this paper can be handled vigilantly by adopting following safety techniques:

### 3.3.1 Measures against JPGvirus

**A. Facebook's Network Security System**

Facebook uses some inbuilt checks for blocking such malicious scripts. If some profile is caught posting such type of links, it is warned for deactivation by facebook security team.

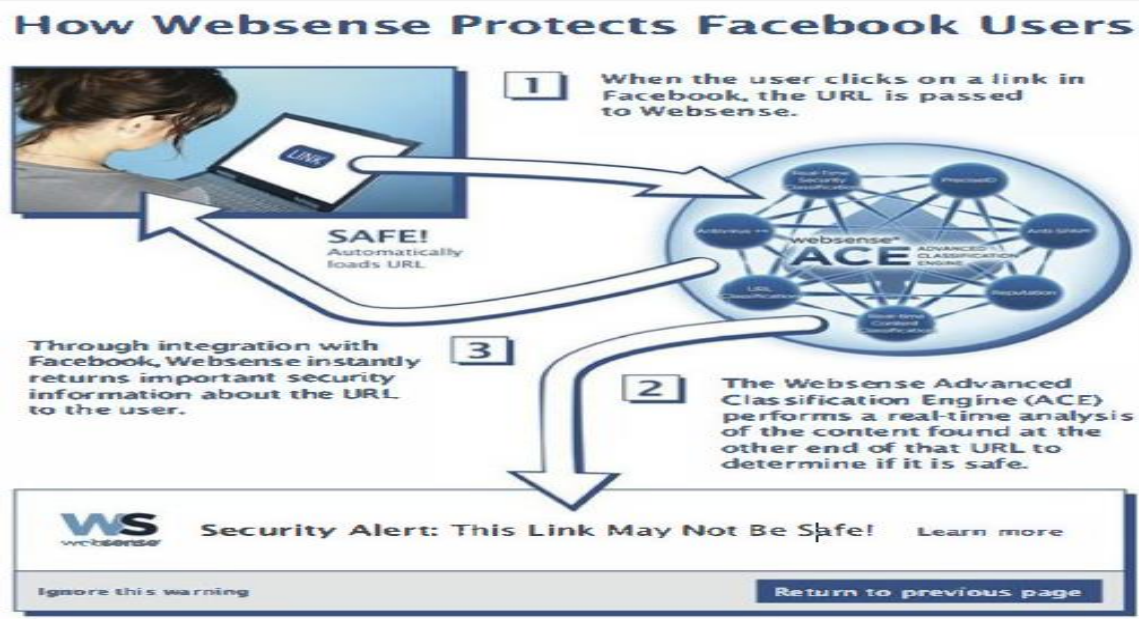

**Figure 5:** Steps for security check by Websense for a facebook internal link

**B. Websense Security Checks**

Websense is a popular computer security firm with which facebook has negotiated so that whenever a link is posted or clicked within facebook, it is examined by Websense ThreatSeeker Cloud and if it comes from an untrusted source, a warning message is shown to the user.

But due to the dynamic nature of the web content, Websense can only check links after posting; it cannot stop them from being posted. This flow can be understood from the figure[17] that is shown above.

### 3.3.2 Measures against Stuxnet Worm

Some possible precautions to avoid the Stuxnet vulnerabilities from propagating are:

**A.** *MS10_061_spoolss vulnerability* can be prevented by not allowing any executable file to be printed by printer. And if any executable file is allowed to be printed, it should have filename less than 10 characters.

**B.** *MS10_046_dllloader vulnerability* can be prevented by not creating shortcut file of internet explorer.

## 4. CONCLUSION

The number of social networking users is increasing day by day all over the world which makes these OSNs more susceptible to the reverse social engineering attacks, one of which is done with the help of JPGvirus discussed in this paper. We surveyed on how the attackers can take advantage of the facebook feature that is, third-party web servers are used to host the facebook applications. If the facebook applications will be hosted from the authentic facebook's servers, then these JPGvirus attacks will lose much of their power.

Another type of worm which was considered here is Stuxnet worm, which was mainly designed to cause the malfunctioning of nuclear power plants. It was designed so well that the system administrator was unaware till its last stage when most of the intended damage was already done and its disastrous effects could not be reversed.

Finally, we aim to provide an understanding about how social network users can be cautious against the reverse social engineering attacks by understanding how a JPGvirus works and as far as Stuxnet worm is concerned above measures can help in maintaining computer system safety to a large extent.

## Authors

Dharmendra Singh has completed his bachelor's in technology from Stani Memorial College of Engineering and Technology Jaipur. He is currently pursuing M.Tech from Government Engineering College, Ajmer. His research interests include Network Security and Cyber warfare.

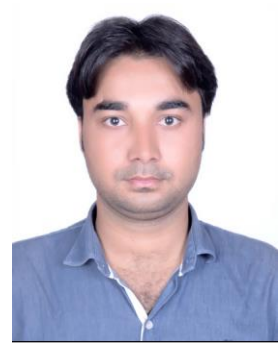

Rakhi Sinha has obtained her bachelor's degree in technology from Govt. Women's Engineering College, Ajmer. She is currently pursuing her master's from Govt. Engineering College, Ajmer. Her research interests include Cyber Espionage and Sabotage.

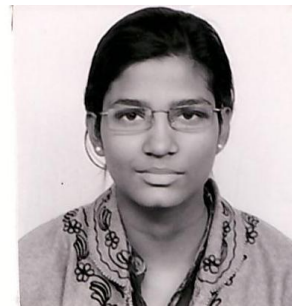

Dr. Rakesh Rathi is currently serving as Assistant Professor in the Department of Computer Engineering in Government Engineering College, Ajmer. He is a CISCO certified academy instructor and is a member of several professional societies. He has taught in various institutions for about 13 years with 6 years of post graduate teaching. He has co-authored and supervised many international and national publications in conferences and journals.

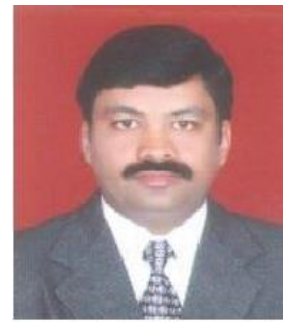